\documentclass[aps,pre,preprint,groupedaddress]{revtex4-1}
\usepackage[T1]{fontenc}
\usepackage[latin9]{inputenc}
\usepackage{amssymb}
\usepackage{graphicx}
\usepackage{epstopdf}
\usepackage{bm}
\usepackage{amsmath}
\DeclareMathOperator{\sign}{sign}
\usepackage{subcaption}
\usepackage{caption}

\def\be{\begin{equation}}
\def\ee{\end{equation}}
\def\bea{\begin{eqnarray}}
\def\eea{\end{eqnarray}}

\begin{document}

\title{Producing Virtually Defect Free Nanoscale Ripples by Ion Bombardment of Rocked Solid Surfaces}

\author{Matt P. Harrison and R. Mark Bradley}
\affiliation{Department of Physics, Colorado State University, Fort
Collins, CO 80523, USA}

\date{\today}

\begin{abstract}
Bombardment of a solid surface with a broad, obliquely-incident ion beam frequently produces nanoscale surface ripples.  The primary obstacle that prevents the adoption of ion bombardment as a nano-fabrication tool is the high density of defects in the patterns that are typically formed.  Our simulations indicate that ion bombardment can produce nearly defect free ripples on the surface of an elemental solid if the sample is concurrently and periodically rocked about an axis orthogonal to the surface normal and the incident beam direction.   We also investigate the conditions necessary for rocking to produce highly ordered ripples and discuss how the results of our simulations can be reproduced experimentally.

\end{abstract}

\pacs{81.16.Rf,79.20.Rf,68.35.Ct}

\date{\today}

\maketitle

\textbf{Introduction} ~ The nanoscale patterns formed by bombardment of a solid surface with a broad beam of noble gas ions have been a subject of interest for decades \cite{MunozGarcia:MaterialsScienceAndEngineeringRReports:2014}.  Ion bombardment has the potential to be an extremely useful and economical way of producing patterns which have a characteristic length of tens of nanometers and which are well ordered over much longer distance \cite{MunozGarcia:MaterialsScienceAndEngineeringRReports:2014}.  A longstanding issue in this field, however, is the high density of defects in the patterns that typically form.  This problem is the primary obstacle that prevents widespread use of ion bombardment as a nanofabrication tool, and much work has been done toward the goal of producing very well ordered patterns \cite{ziberi2005ripple,cuenat2005lateral,Ziberi:AppliedPhysicsLetters:2008,
Keller:PhysicalReviewB:2010,motta2012highly,PhysRevB.86.121406,Mollick14,Ou15}.
To date, no experiment has yielded highly ordered patterns on an elemental sample using a noble gas ion beam.

In this rapid communication, we present the results of numerical simulations of an elemental surface that is bombarded by a broad beam of noble gas ions with a polar angle of incidence that varies periodically in time.  In an experiment, this could be achieved by rocking the sample about an axis orthogonal to the surface normal and the incident ion beam. We take the equation of motion in the absence of rocking to be the usual anisotropic Kuramoto-Sivashinsky (AKS) equation \cite{PhysRevLett.74.4746,Makeev:NuclearInstrumentsAndMethodsInPhysicsResearch:2002,MunozGarcia:MaterialsScienceAndEngineeringRReports:2014}. Several of the coefficients in this equation depend on the angle of incidence, and so periodic sample rocking has the effect of making these coefficients periodic in time.  We show that a remarkable and unforeseen degree of order emerges for a broad range of parameters in both one-dimensional (1D) and two-dimensional (2D) simulations.  Thus, periodic temporal driving can lead to near perfect spatial periodicity. 

\textbf{1D Results}  ~The equation of motion for the unrocked solid surface is the much-studied AKS equation,  \cite{Kuramoto, Sivashinsky,michelson1986steady,Rost95}
\be
u_t=v_0' u_x-A u_{xx} + A' u_{yy} - B\nabla^4 u + \left(\lambda u_x^2+\lambda' u_y^2\right)/2\label{AKS},
\ee
where $u$ is the deviation of the surface height from its unperturbed steady-state value; $x$ and $y$ are the horizontal coordinates parallel and perpendicular to the projection of the ion beam direction onto the surface, respectively; $t$ is the time; and the subscripts $x$, $y$, and $t$ denote partial derivatives. All of the constant coefficients $A$, $A'$, $\lambda$, $\lambda'$ and $B$ will be taken to be positive save for $\lambda$.  The AKS equation has long been used as a model for the time evolution of an ion bombarded surface \cite{PhysRevLett.74.4746,Makeev:NuclearInstrumentsAndMethodsInPhysicsResearch:2002}. and expressions that relate the coefficients to the underlying physical parameters have been given \cite{bradley1988theory,Carter:PhysRevBCondensMatter:1996,Makeev:NuclearInstrumentsAndMethodsInPhysicsResearch:2002}  . The term proportional to the constant $v_0'$ in Eq.~(\ref{AKS}) may be eliminated by transforming to a moving frame of reference, and so will be dropped for the remainder of this paper.  

In this section, we specialize to the case in which $h$ is independent of $y$.  This reduces Eq.~(\ref{AKS}) to the 1+1 dimensional Kuramoto-Sivashinsky (KS) equation,
\be
u_t=-A u_{xx}-Bu_{xxxx}+\lambda u_x^2/2\label{KS}.
\ee
For the case of time-independent coefficients, the KS equation can be rescaled to a completely parameter-free form by setting 
\be
u=  (A/\lambda)~\tilde u,~~~
t= (B/A^2)~\tilde t,~~~\textrm{and}~~
x= \sqrt{B/A}~\tilde x,\label{rescale}
\ee
where $\tilde u$, $\tilde t$, and $\tilde x$ are the dimensionless surface height, time, and lateral coordinate, respectively.  

To investigate the effects of periodic rocking on the pattern formation, we focus on the special case in which the sample is bombarded for a time $t=\pi/\omega$ at an angle of incidence $\theta_1$, then for an equal time at an angle of incidence $\theta_2$, and so forth.  Without loss of generality, we may express the values of $A$ and $\lambda$ at these angles as
\bea
&&\lambda(\theta_1)=\lambda_0~(1+r_1), ~~~~~~~~~~ A(\theta_1)=A_0~(1+r_2),\label{eq1}\\
&&\lambda(\theta_2)=\lambda_0~(1-r_1), ~~~\text{and}~~  A(\theta_2)=A_0,\label{eq2}
\eea
where $r_1>0$ and $r_2$ are dimensionless parameters and $\lambda_0$ and $A_0$ are constants.  While $A(\theta)$ and $\lambda(\theta)$ are not simple functions of the angle of incidence $\theta$, because we are switching the angle of incidence discretely, we only need the values of $A$ and $\lambda$ at two angles.  

Because neither $\lambda(\theta)$ nor $A(\theta)$ is a monotone function  \cite{Makeev:NuclearInstrumentsAndMethodsInPhysicsResearch:2002} of $\theta$, it is possible to choose values of $\theta$ for which $A(\theta_1)=A(\theta_2)$ and $\lambda(\theta_1)\neq\lambda(\theta_2)$.  Thus, by a suitable experimental setup, a periodic, discrete variation in $\lambda$ can be achieved while minimizing or eliminating any variation in $A$.  As shown below, oscillations in $A$ can be detrimental to the formation of highly ordered patterns, and should be minimized when possible.  We will not consider the effect of periodic variation in $B$, since its dependence on the angle of incidence is weak \cite{MunozGarcia:MaterialsScienceAndEngineeringRReports:2014}.

\begin{figure}[h]
\begin{centering}
\includegraphics[width=3.0in]{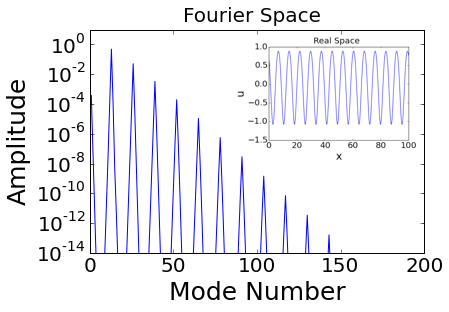}
\captionsetup{justification=raggedright,singlelinecheck=false}

\caption{(Color online) Plots of the rocked surface in real space (inset) and in Fourier space for $r_1=2$, $r_2=0$, dimensionless frequency $\tilde \omega\equiv\omega B/A_0^2=0.15\pi$, domain length $L=100$, and $\tilde t=10^3$.}\label{CTP}
\par\end{centering}
\end{figure}

\begin{figure}[h]
\begin{centering}
\includegraphics[width=3.0in]{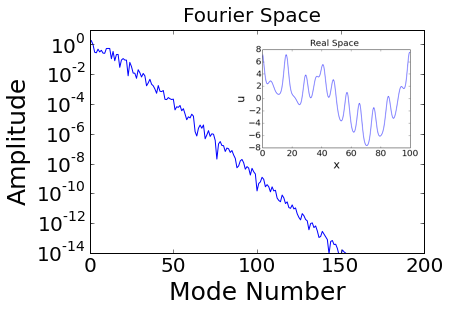}
\captionsetup{justification=raggedright,singlelinecheck=false}
\caption{(Color online) Plots of the unrocked surface in real space (inset) and in Fourier space for $r_1=r_2=0$, $L=100$ and $\tilde t=10^3$.}\label{NTP}
\par\end{centering}
\end{figure}

\begin{figure}[h]
\begin{centering}
\includegraphics[width=3.0in]{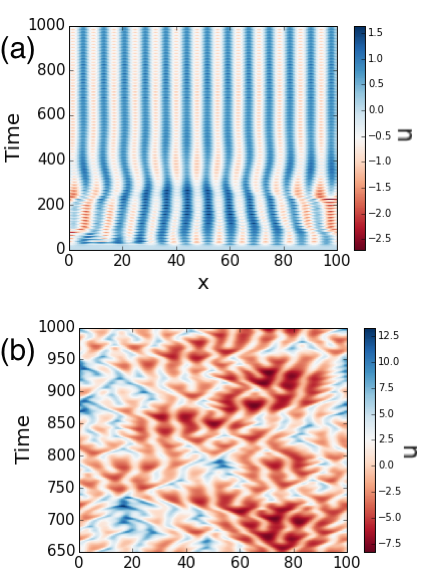}
\captionsetup{justification=raggedright,singlelinecheck=false}
\caption{(Color online) Spacetime plots for (a) $r_1=2$ (rocked), and (b) $r_1=0$ (unrocked) respectively with $\tilde \omega=0.15 \pi$ and $r_2=0$.  A shorter time scale was used for the unrocked case so that the finer structure is visible.}\label{histplots}
\par\end{centering}
\end{figure}

Using exponential time differencing \cite{cox2002exponential,kassam2005fourth}, we have performed numerical integrations of Eq.~(\ref{KS}) with $\lambda$ and $A$ periodically and discretely switching between the values given by Eqs.~(\ref{eq1}) and (\ref{eq2}).  We employed the rescaling given by Eq.~(\ref{rescale}) with $A$ and $\lambda$ replaced by $A_0$ and $\lambda_0$, respectively. The initial condition was low amplitude spatial white noise.  The results are nothing short of astonishing.  Figures~\ref{CTP} and \ref{NTP} show the results of two simulations, one with rocking and one without, in real space and Fourier space.  While the KS equation without rocking yields a surface which has a high degree of disorder, albeit with a characteristic length scale, the rocked KS equation with
dimensionless frequency $\tilde \omega\equiv\omega B/A_0^2=0.15\pi$
produces ripples that are almost perfectly periodic.  Figure~\ref{histplots} shows spacetime plots of time sequences taken from the same simulations.  These demonstrate that without rocking the system exhibits the spatio-temporal chaos characteristic of the KS equation \cite{Kuramoto, Sivashinsky}.  On the other hand, following a brief transient state, the rocked sample displays an extremely high degree of order which persists over time.  Larger domain sizes $L$, finer spatial and temporal discretizations, and much longer simulation times have been investigated numerically, and give comparable results to those shown in Figs.~1-3.

The oscillations in ripple amplitude evident in Fig.~\ref{histplots}(a) may be explained by a scaling argument.  As discussed above, if $A$ and $\lambda$ are time-independent constants, then Eq.~(\ref{KS}) may be written in a parameter-free form via the rescaling given by Eq.~(\ref{rescale}).  This shows that the characteristic ripple amplitude is proportional to $|A/\lambda|$.  For the case in which the sample is rocked and $r_2=0$, we therefore expect that after a sudden change in the value of $\lambda$, the amplitude of the ripple pattern will evolve toward the amplitude associated with the new value of $\lambda$.  When $|\lambda|$ decreases the ripples grow larger in amplitude, and when $|\lambda|$ increases the amplitude attenuates. 

An intuitive understanding of the order produced by rocking may be gained by a heuristic argument.  Consider what happens to the surface when $|\lambda|$ changes from a large value to a small value at a time $t_1$.  In the moments before the switch, the term proportional to $\lambda u_x^2$ in the rocked KS equation is on average comparable to the linear terms $-Au_{xx}-Bu_{xxxx}$ since in the steady-state limit, the ripple growth rate averages to zero.  Immediately after the switch, therefore, the term $\lambda u_x^2$ will typically be small compared to the linear terms.  Consequently, we expect the Fourier transform of the surface height $\tilde{U}(k,t)$ to grow approximately as
\be
\tilde{U}(k,t)\simeq\tilde{U}(k,t_1)\exp\left((Ak^2-Bk^4)~ (t-t_1)\right)\label{lineom},
\ee
where $k$ is the wave number.  Thus, periodically reducing the value of $|\lambda|$ allows the surface to periodically grow roughly as it would in the linear approximation.  Ripples described by Eq.~(\ref{lineom}) become increasingly well ordered because the peak in the Fourier spectrum becomes higher and narrower as time passes.  Conversely, when the value of $|\lambda|$ is increased, the amplitude gained during the stage of approximately linear growth attenuates according to the scaling argument given above.  This ensures that the ripple amplitude does not become so large that higher order nonlinear effects would have to be taken into account.
This explanation suggests that good order will not obtained if $\lambda$ oscillates about an average value of zero, since then there is no opportunity for nearly linear growth to take place.  

The excitation of Fourier modes which are multiples of the selected wave number seen in Fig.~\ref{CTP} arises as a consequence of the coupling between modes induced by the nonlinear term $\lambda u_x^2$.  A large amplitude mode with wavenumber $k$ will directly excite the mode of wavenumber $2k$.  The coupling between these two modes will then excite the mode of wavenumber $3k$, and so on.

To characterize the quality of the order produced by the rocking procedure, we fit the peak surrounding the highest amplitude wavenumber in the Fourier spectrum to a Gaussian and record its width.  In order to avoid sampling at the same point in each rocking cycle, the fits were performed at hundreds of randomly selected times throughout a given simulation and then averaged. 

\begin{figure}[h]
\begin{centering}
\includegraphics[width=3.0in]{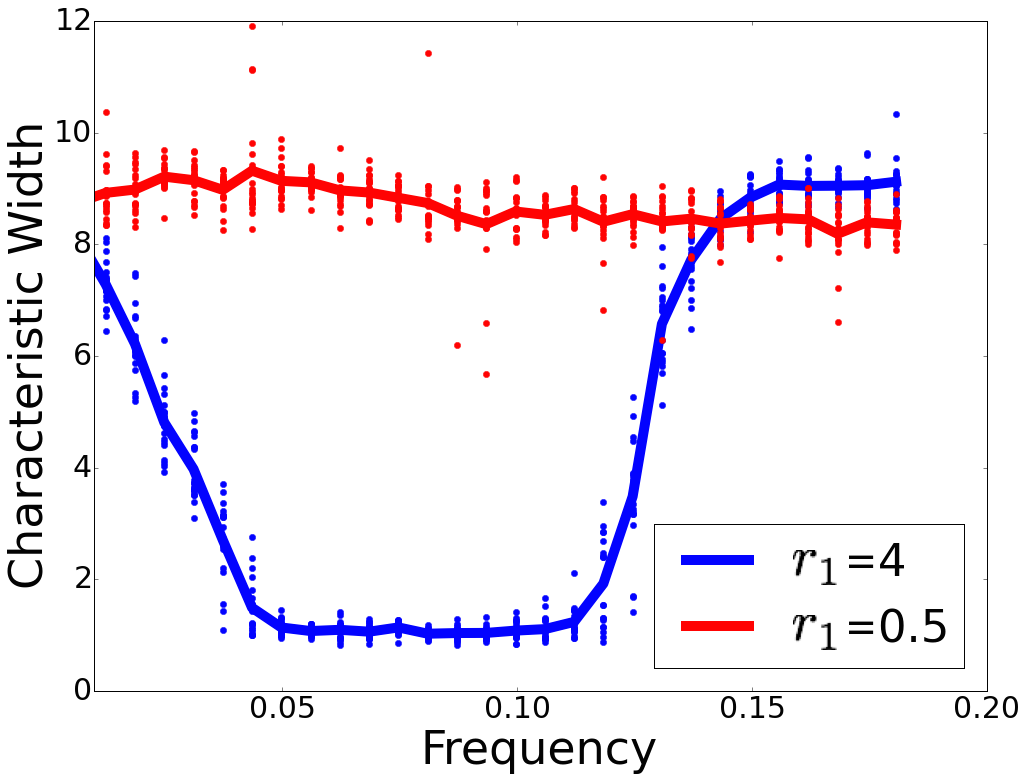}
\captionsetup{justification=raggedright,singlelinecheck=false}
\caption{(Color online) Fourier peak width as a function of the scaled rocking frequency $\tilde f\equiv\tilde\omega/(2\pi)$ for two values of $r_1$ and with $r_2=0$.  Each point represents a single simulation, and the average for each value of the frequency is shown.  Note that for $r_1=0.5$, $\lambda$ always has the same sign, while for $r_1=4$, it changes sign.}\label{wvomega}
\par\end{centering}
\end{figure}

 Figure~\ref{wvomega} shows the width of the highest peak in the Fourier spectrum as a function of the rocking frequency for two values of $r_1$.  It is clear that the effect of rocking is strongly dependent on the frequency with which the sample is rocked.  However, within a broad range of frequencies, the surface becomes highly ordered for $r_1=4$, making this procedure feasible to implement experimentally.  One important conclusion drawn from our simulations is that, just as our heuristic argument suggested, it is essential that $r_1>1$ (so that $\lambda$ changes sign periodically) for good order to form.  This is also illustrated by Fig.~\ref{wvomega}, since good order is not obtained for $r_1=0.5$.  Increasing $r_1$ further than $r_1=4$ has the effect of slightly narrowing the band of frequencies which produce good order and reducing the amplitude of the resulting ripples.  Nevertheless, for $1.5\lesssim r_1\lesssim 15$, scaled frequencies $\tilde f \equiv \tilde \omega/2 \pi$ between $0.06$ and $0.08$ produce exceptionally good order.

\begin{figure}[h]
\begin{centering}
\includegraphics[width=3.0in]{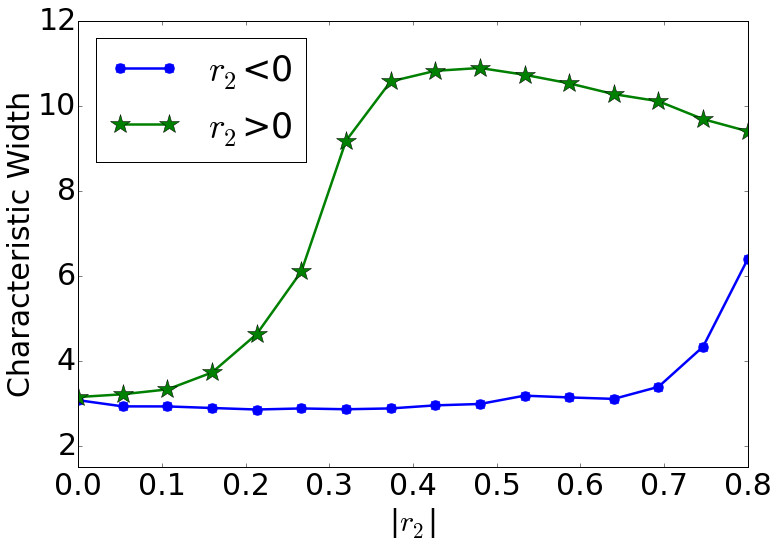}
\captionsetup{justification=raggedright,singlelinecheck=false}
\caption{(Color online) Fourier peak width for $r_1=4$ and $\tilde\omega=0.15\pi$ for opposite signs of $r_2$.  For $r_2<0$ the surface continues to form nearly perfect ripples for relatively large values of $|r_2|$.}\label{wva}
\par\end{centering}
\end{figure}

An important physical consideration is the effect that a periodic variation of the coefficient $A$ in Eq.~(\ref{KS}) has on the order obtained by rocking.  Since this coefficient depends on the angle of incidence, it is likely to vary in general unless $\theta_1$ and $\theta_2$ are carefully selected.  For $\lambda$ and $A$ given by Eqs.~(\ref{eq1}) and (\ref{eq2}), the characteristic width of the highest Fourier peak for $r_1=4$ and a range of $r_2$ values is shown in Fig.~\ref{wva}.  For small positive values of $r_2$, the surface still becomes well ordered, but larger positive values of $r_2$ do not result in a well ordered surface.  If $r_2<0$, on the other hand, the high degree of order develops even for relatively large values of $|r_2|$.  Thus, the variation in $A$ due to the rocking procedure is not expected to be a significant impediment to producing virtually defect free ripples by sample rocking.

 If the order seen in our simulations is to be achieved experimentally, it is crucial that the rocking  frequency be chosen within the frequency range discussed above.  Fortunately, finding the correct rocking frequency only requires that the linear growth rate be determined for one of the two angles of incidence.  The dimensional and nondimensional rocking frequencies are related by $f=4\sigma\tilde f$, where $\sigma$ is the linear growth rate of the fastest growing mode for that angle of incidence.  Experimentally, $\sigma$ may be approximated by the rate at which the surface roughens at early times, since this roughening will be dominated by the most linearly unstable mode. Simulations indicate that the growth rate during the period where $|\lambda|$ is a minimum determines the optimal rocking frequency.  Therefore, if $r_1$ is greater (less) than zero, then $\sigma$ should be evaluated at $\theta_2$ ($\theta_1$).

As we have seen, for the rocking procedure to be effective, $\lambda$ must change sign.  There is strong theoretical and experimental evidence that $\lambda$ passes through zero at a critical angle $\theta_c$ for unrocked samples  \cite{Makeev:NuclearInstrumentsAndMethodsInPhysicsResearch:2002,Engler14}.  Typically, $60^\circ<\theta_c<80^\circ$.  Given that the amplitude of the surface roughness scales as $|\lambda|^{-1}$, we expect that $\lambda$ will vanish at the angle of incidence which maximizes the surface roughness.  Note that while solutions to the KS equation grow without limit for a spatial white noise initial condition when $\lambda=0$, the amplitude remains finite in an experiment because a finite ion fluence is used.

\begin{figure}[h]
\begin{centering}
\includegraphics[height=2.4in]{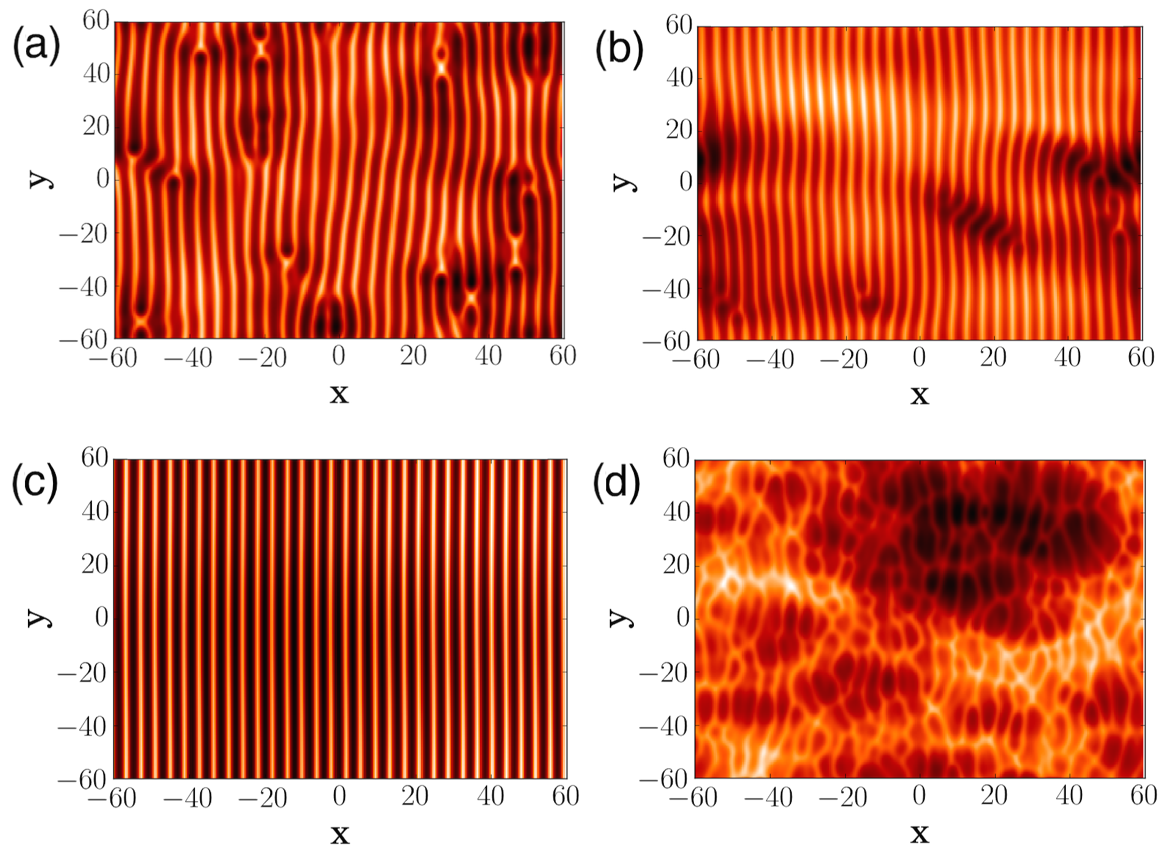}
\captionsetup{justification=raggedright,singlelinecheck=false}
\caption{(Color online) (Multimedia View) (a) - (c) A time series for a square domain of side length $L=120$ with $A=A'=1$, $B=1$, $\lambda=2+8\sign(\sin(
\tilde\omega \tilde t ))$, $\lambda'=2$, and $\tilde\omega=0.15\pi$ at times $\tilde t=106,330$, and $1840$.
(d) A simulation with the same parameters as in (a) - (c) but with $\lambda=2$ so that there is no rocking.  The time is $\tilde t=1840$. }\label{2Dpics}
\par\end{centering}
\end{figure}

\textbf{2D Results}    ~Given the degree of order that can form on a rocked surface in 1D, it is natural to ask whether a 2D rocked surface will produce similar results.  Physically, this means we are no longer requiring that $h$ be independent of $y$.  We return to Eq.~(\ref{AKS}), keeping $v_0'=0$.  With temporally periodic coefficients, Eq.~(\ref{AKS}) is the rocked AKS equation.  Panels (a)-(c) of Fig.~\ref{2Dpics}  show the surface for a particular set of parameters for which $\lambda$ oscillates between the values 10 and $-6$.  As in 1D, the unrocked equation of motion displays spatio-temporal chaos with a characteristic length scale (see Fig.~\ref{2Dpics} (d)).  The rocked AKS equation, on the other hand, initially forms a transient state which contains numerous defects.  The rocking procedure causes these defects to move together and annihilate. Eventually, even long wavelength Fourier modes are suppressed. The full video from which these snapshots are taken is available in the link in Fig.~(\ref{2Dpics}) (Multimedia View).

The rocked AKS equation has more parameters than the rocked KS equation, and the computational time required for a complete investigation of the parameter space of the rocked AKS equation would be prohibitively long.  However, variation in the coefficient $\lambda'$ is likely unavoidable during rocking, since this coefficient also depends on the angle of incidence $\theta$.  We therefore considered the effect of simultaneously varying $\lambda'$ and $\lambda$.   The results of a simulation for which $\lambda'=2+0.2\sign(\sin( \omega  t ))$ (not shown) revealed that periodic oscillations of $\lambda'$ with this amplitude do not have a detrimental effect on the resultant order for $A$, $A'$, $B$, and $\lambda$ equal to their values in the first three frames of Fig.~\ref{2Dpics}.

\textbf{Discussion}~ The consequences of a time-periodic coefficient in the KS equation have been considered in the context of annular fluid flow \cite{coward1995nonlinear}.  Due to computational limitations at the time of that work, the phenomenon reported here was not discovered.   In the first theoretical study of a periodically rocked, ion bombarded surface \cite{tagg1986sample}, several important physical contributions to the dynamics were neglected, including curvature dependent sputtering \cite{bradley1988theory}.  In a later theoretical treatment of ion bombardment with sample rocking \cite{Carter:ApplPhysLett:1997}, nonlinear terms were omitted from the equation of motion and consequently no increase in order was found.  Experiments have not yet been performed in which a sample was bombarded while being periodically rocked.  Our results give a compelling motivation for conducting experiments of that kind. By contrast, azimuthal sample rotation during ion bombardment has studied intensively, and gives a means of producing ultra-smooth surfaces \cite{zalar1986auger,bradley1996theory,bradley1996dynamic,Frost08}, generating hexagonal order \cite{Frost00}, and controlling ripple patterns \cite{harrison2015nanoscale}.

The case in which $A$ and $\lambda$ vary sinusoidally in time has been explored numerically, and was found to produce order comparable to discrete switching.  More complicated time dependencies are beyond the scope of this letter, but are not expected to produce substantially different results.

\textbf{Conclusion}~Our simulations demonstrate that if a sample is bombarded with a broad noble gas ion beam while simultaneously being rocked, nearly perfect nanoscale ripples can result.  
Unlike other methods \cite{Mollick14,Ou15},
ours can be used to produce highly ordered surface ripples on an elemental material and does not require the implantation of an undesirable second atomic species or a high sample temperature.
We also discussed how optimal values of the rocking frequency and the angles of incidence may be determined in an experiment.

R.M.B. is grateful to the National Science Foundation for its support through grant DMR-1305449.  The authors thank Dan Pearson for stimulating discussions and for sharing his software.
\raggedright

\bibliographystyle{apsrev4-1}

\end{document}